\documentclass{article}
\usepackage{array}
\usepackage{amsmath}
\usepackage{graphicx}
\usepackage{amssymb}
\usepackage[parfill]{parskip}
\usepackage{wrapfig}
\usepackage{color}
\usepackage{caption}
\usepackage{parskip}
\usepackage{ulem}
\usepackage{multicol}
\usepackage[margin=1in]{geometry}
\usepackage[english]{babel}
\usepackage[bottom]{footmisc}
\usepackage[utf8]{inputenc}
\usepackage{fancyhdr}
\usepackage{authblk}

\begin{document}
\title{Tunable colloid trajectories in nematic liquid crystals near wavy walls}

\author[1]{Yimin Luo}
\author[2]{Daniel A. Beller}
\author[1]{Giuseppe Boniello}
\author[3]{Francesca Serra$^*$}
\author[1]{Kathleen J. Stebe$^*$}

\affil[1]{\footnotesize Chemical and Biomolecular Engineering, University of Pennsylvania, Philadelphia, PA 19104, USA}
\affil[2]{Brown University School of Engineering, Providence, RI 02912, USA}
\affil[3]{Physics and Astronomy, Johns Hopkins University, Baltimore, MD 21218, USA}
\affil[*]{Corresponding authors}

\maketitle 
\begin{abstract}
The ability to dictate colloid motion is an important challenge in fields ranging from materials
science to living systems. Here, by embedding energy landscapes in confined nematic liquid crystals, we design a versatile platform to define  colloidal migration. This is achieved by placing a wavy wall, with alternating hills and wells, in nematic liquid crystals, to impose a smooth elastic energy field with alternating splay and bend distortions. This domain generates (meta) stable loci that act as attractors and unstable loci that repel colloids over distances large compared to the colloid radius. Energy gradients in the vicinity of these loci are exploited to dictate colloid trajectories.
We demonstrate several aspects of this control, by studying transitions between defect configurations, propelling particles along well defined paths and exploiting multistable systems to send particles to particular sites within the domain. Such tailored
landscapes have promise in reconfigurable systems and in microrobotics applications. \end{abstract}

\begin{multicols}{2}

Ever since Brown discovered the motion of inanimate pollen grains, material scientists have been fascinated by the vivid, life-like motion of colloidal particles. Indeed, the study of colloidal interactions has led to the discovery of new physics and has fueled the design of functional materials \cite{manoharan2015colloidal, dinsmore1998self, sacanna2013}. External applied fields provide important additional degrees of freedom, and allow microparticles to be moved along energy gradients with exquisite control. In this context, nematic liquid crystals (NLCs) provide unique opportunities \cite{blanc2013ordering}. Within these fluids, rod-like molecules co-orient, defining the nematic director field \cite{DeGennes}. Gradients in the director field are energetically costly; by deliberately imposing such gradients, elastic energy fields can be defined to control colloid motion. Since NLCs are sensitive to the anchoring conditions on bounding surfaces \cite{anchoring, brake2003effect}, reorient in electro-magnetic fields \cite{brochard1972dynamics, DeGennes}, have temperature-dependent elastic constants \cite{DeGennes} and can be reoriented under illumination using optically-active dopants \cite{matczyszyn2003phase, legge1992azo}, such energy landscapes can be imposed and reconfigured by a number of routes.

Geometry, topology, confinement and surface anchoring provide versatile means to craft elastic energy landscapes \cite{chen2018colloidal, terentjev1995disclination,yoshida2015three, serra2016}. By tailoring bounding vessel shape and NLC orientation at surfaces, elastic fields to direct colloid assembly can be defined \cite{blanc2013ordering}. This was shown for NLC controlled by patterned substrates \cite{li2017directed, Peng2016}, optically manipulated in a thin cell \cite{nych2013assembly}, or in micropost arrays \cite{cavallaro2013exploiting, Lee2016}, grooves \cite{luo2016around, sengupta2013, fukuda2012}, and near wavy walls \cite{luo2016experimental, silvestre2004key}. In prior work, the energy fields near wavy walls have been exploited to demonstrate lock-and-key interactions, in which a colloid (the key) was attracted to a particular location (the lock) along the wavy wall to minimize distortion in the nematic director field. However, the elastic energy landscapes obtainable with a wavy wall are far richer, and provide important opportunities to direct colloidal motion that go far beyond near-wall lock-and-key interaction (Fig. 1). Fig. 1a shows sample trajectories that a particle can take next to a wall thanks to the rich energy landscape. In this system, elastic energy gradients are defined by the period and amplitude of the wavy structure (Fig. 1b-c), allowing long ranged wall-colloid interactions. Colloids can be placed  at  equilibrium sites far from the wall that can be tuned by varying wall curvature. Unstable loci, embedded in the elastic energy landscape, can repel colloids and drive them along multiple paths. In this work, we develop and exploit aspects of this energy landscape to control colloid motion. For example, we exploit metastable equilibria of colloids to induce gentle transformations of the colloids' companion topological defects driven by the elastic fields (Fig. 1d). Since topological defects can be sites for accumulation of nanoparticles and molecules, such transformations will enable manipulation of hierarchical structures.  We also create unstable loci to direct particle trajectories and to produce multistable systems, with broad potential implications for reconfigurable systems and microrobotics (Fig. 1e-g).

In NLCs, topology, confinement, and surface anchoring dictate colloid interactions with elastic energy landscapes. This well-known behavior \cite{poulin1997p, blanc2013ordering} implies that strategies to dictate colloidal physics developed in these systems are robust and broadly applicable to any material with similar surface anchoring and shape. Furthermore, the ability to control the types of topological defects that accompany colloidal particles provides access to significantly different equilibrium states in the same system. Isolated colloids with homeotropic (perpendicular) anchoring are accompanied by a topologically required companion defect \cite{terentjev1995disclination,kleman1989defects}. This defect adopts either a hyperbolic hedgehog configuration, with a topological point defect near the particle, or a Saturn ring configuration with a disclination line encircling the particle. The hedgehog configuration is commonly observed for micron-sized particles, although the Saturn ring can be stabilized by confinement \cite{gu2000observation}. Similarly, a colloid with planar anchoring forms two topologically required ``boojums", surface defects at opposing poles \cite{poulin1998inverted}. Together, the hedgehog and the colloid form a topological dipole, while colloids with Saturn ring or boojums companion defects have quadrupolar symmetries. Far from the colloid, the corresponding disturbances in the director field are analogous to the electrostatic potential distribution around charged multipoles \cite{lubensky1998topological, musevic2008, Lapointe2009, senyuk2016hexadecapolar}. Thus, the structure of the colloid and its companion defect dictate the range and form of their interactions. In this research, we demonstrate control over each of these defect configurations, focusing initially on colloids in the Saturn ring configuration. \\

\begin{figure*}[htb]
 \centering
  \includegraphics[scale=0.35]{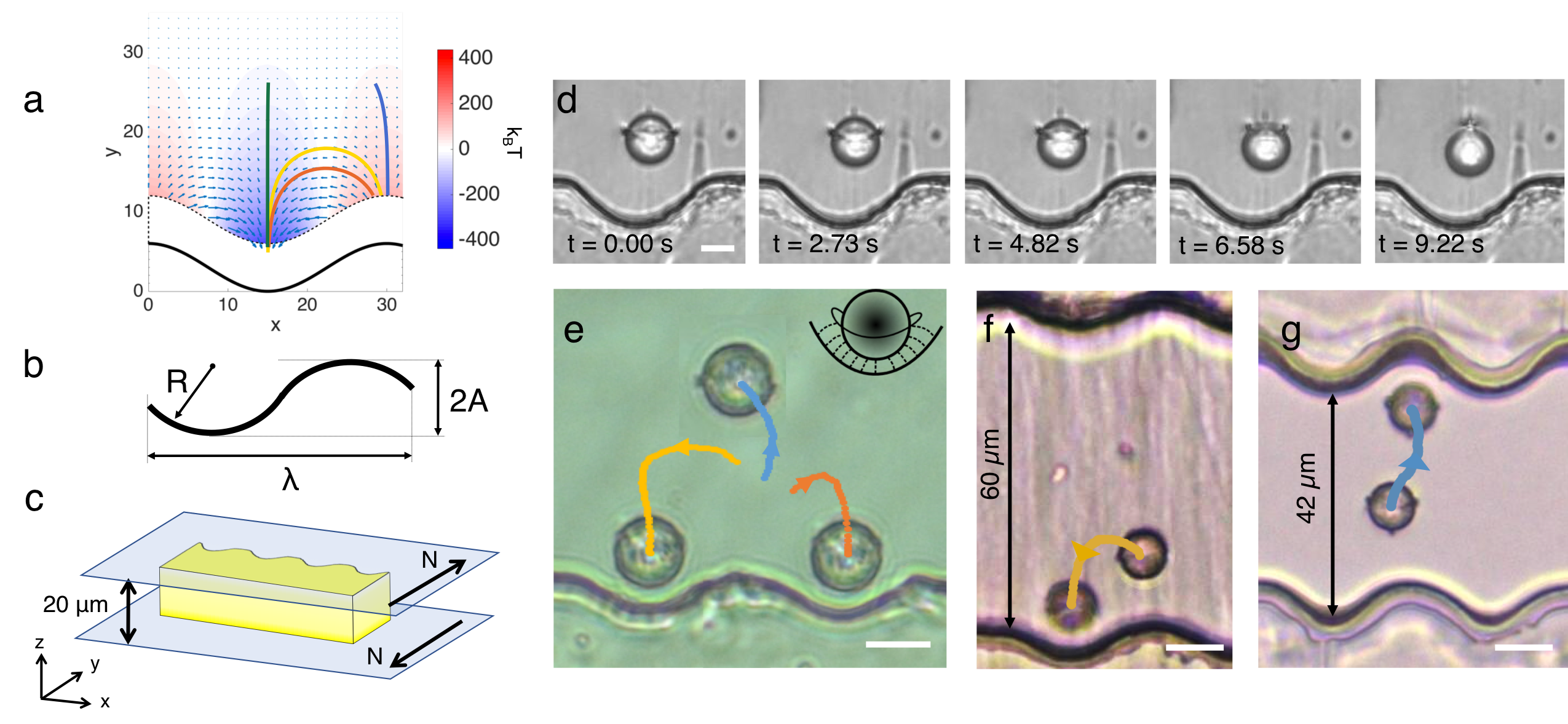}\\
 \caption{{\bf Energy landscape for versatile path-plannning.}  (a) A heat map of the elastic energy for a colloid in the vicinity of the wall as calculated by LdG simulation obtained by placing the center of mass (COM) of a colloid at different locations of ($x$, $y$), with reference energy evaluated at ($\lambda$/2, $\lambda$). The energy in the color bar is in $k_B$T. The vector field on this figure shows local elastic forces on the particle, obtained from the gradient in the elastic energy field. The solid curves indicate a few predicted trajectories for colloids placed at different initial positions in the energy landscape. Further details of how this energy landscape is generated can be found in SI and Fig. S2. (b) Schematics of the wall shape with relevant parameters. (c) Schematic of the experimental setup (N denotes rubbing direction). (d) Time-lapse image of a Saturn ring transforming to a dipole at a metastable position remote from the wall defined by the elastic energy field. (e) The colored dots represent three trajectories traced by a homeotropic colloid released from an initial position between two identical wells. With small changes in the initial position, the colloids end up in three possible locations, all shown in the figure. Inset: A sketch of the director field around a homeotropic colloid docked inside of a well. (f,g) A magnetic particle with a Saturn ring defect, placed near a hill, with wall to wall separation in (f) 60 $\mu$m and (g) 42 $\mu$m. In (f), the particle is more attracted to the wall on the same side. In (g) the particle is moved away from the wall by repulsion from the hill, and traverses the separation between walls to land in the opposite side (2a = 9 $\mu$m). The scale bars are 10 $\mu$m.}
\end{figure*}

\section*{Results}
To mold the elastic energy landscape near a curved boundary (Fig. 1a), we fabricate long, epoxy resin strips using standard lithographic techniques to form wavy structures (Fig. 1b) that are placed between two parallel glass slides with planar anchoring oriented perpendicular to the strip (see Methods for details, Fig. 1c). This cell is filled by capillarity with a suspension of colloids in the NLC 4-cyano-4'-pentylbiphenyl (5CB) in the isotropic phase, and subsequently quenched into the nematic phase ($T_{NI} = 34.9 \,^o C$). Colloid migration within this assembly is observed with an optical microscope. For the larger beads, as expected, strong confinement between the glass slides stabilizes the Saturn ring configuration \cite{gu2000observation}. Smaller beads, which experience weaker confinement, adopt the dipolar structure. Particles are equally repelled by elastic interactions with the top and bottom glass slides, whose strength dominates over the particles' weight, so gravity plays a negligible role in our system. When observed through the microscope, this configuration forms a quasi-2D system in the ($x,y$) plane, where  $y$ is measured from the base of a well in the direction perpendicular to the wall. The wavy wall forms a series of hills and valleys, with a distance 2$A$ from the base of the well to the highest point on a hill. Because of strong homeotropic anchoring at the wavy wall, these features impose zones of splay and bend in this domain. In particular, the valleys are sites of converging splay, the hills are sites of diverging splay, and the inflection points are sites of maximum bend (see Fig.\ S1 for a detailed discussion on the geometry of the well and the director field). In terms of the parameters $R$ and $A$, the structure has period $\lambda=4R\sqrt{\frac{A}{R}(2-\frac{A}{R})}$. Throughout this study, unless specified otherwise, 2$A$ = 10 $\mu$m.  The gentle undulations of this wall deform the surrounding director field but do not seed defect structures into the NLC. We characterize the control we achieve over colloidal motion by characterizing particle behavior within the energy landscape near this wall. In addition, we use Landau-de Gennes (LdG) simulations of the liquid crystal orientation to guide our thinking. Details of the simulation can be found in the Methods section.

{\bf Attraction to the wall.} 
To determine the range of interaction of a colloid with undulated walls of differing $\lambda$, a magnetic field is used to move a ferromagnetic colloid ($a$ = 4.5 $\mu$m) to a position $y$ far from the wall and $x$ corresponding to the center of the well. The magnet is rapidly withdrawn, and the colloid is observed for a period of 2 min. If the colloid fails to approach the wall by distances comparable to the particle radius within this time, the colloid is moved closer to the wall in increments of roughly a particle radius until it begins to approach the wall. We define the range of interaction $H^*$ as the maximum distance from the wall at which the colloid starts moving under the influence of the wall (Fig. 2). In these experiments, the Saturn ring defect was sometimes pinned to the rough surface of the ferromagnetic particles. To eliminate this effect, these experiments were repeated with homeotropic magnetic droplets with a smooth interface whose fabrication is described in the Methods section. The results with the two systems are similar. A typical trajectory is shown in Fig. 2a in equal time step images ($\Delta t$ = 125 s). For small $\lambda$ (i.e. $\lambda$ $\lesssim 40 \, \mu m$), $H^*$ increases roughly linearly with $\lambda$. However, at larger $\lambda$, the range of interaction increases only weakly. A simple calculation for the director field near a wavy wall in an unbounded medium in the one constant approximation and assuming small slopes predicts that the distortions from the wall decay over distances comparable to $\lambda$ \cite{luo2016experimental}. However, when $\lambda \gg T$, thickness of the cell, confinement by the top and bottom slides truncates this range (See SI and Fig. S3), giving rise to the two regimes reported in Fig. 2b: one that complies with the linear trend and one that deviates from it. A similar shielding effect of confinement in a thin cell was reported in the measurements of interparticle potential for colloids in a sandwich cell \cite{vilfan2008confinement}.

The colloid moves toward the wall along a deterministic trajectory. Furthermore, it moves faster as it nears the wall (Fig. 2c), indicating steep local changes in the elastic energy landscape. This motion occurs in creeping flow (Reynolds number $\mathrm{Re} = \rho v a/\eta \approx 1.15 \times 10^{-8}$, where $\rho$ and $\eta$ are the density and viscosity of 5CB, respectively). The energy dissipated to viscous drag along a trajectory $U$ can be used to infer the total elastic energy change; we perform this intergration and find  $U \sim 5000\, k_{B}T$. In this calculation, we correct the drag coefficient for proximity to the wavy wall according to \cite{brenner1961slow} and for confinement between parallel plates according to \cite{ganatos1980strong} (see \cite{luo2016experimental} for more details). The dissipation calculation shows that gradients are weak far from the wall and steeper in the vicinity of the wall. The elastic energy profile found from the LdG simulation as a function of particle distance from the base of the well is consistent with these observations (Fig. S4). The particle finds an equilibrium position in the well. At larger distances from the wall, the energy increases first steeply, and then tapers off (Fig. S4). For wide wells ($\lambda>$ 15$a$), the energy gradient in x near the wall is weak, and the drag is large. In this setting, the colloid can find various trapped positions, and introduce error to energy calculation. Therefore, the trajectory is truncated at around $y$ = 15 $\mu$m from contact with the wall. 

{ \centering
 \includegraphics[width=\columnwidth]{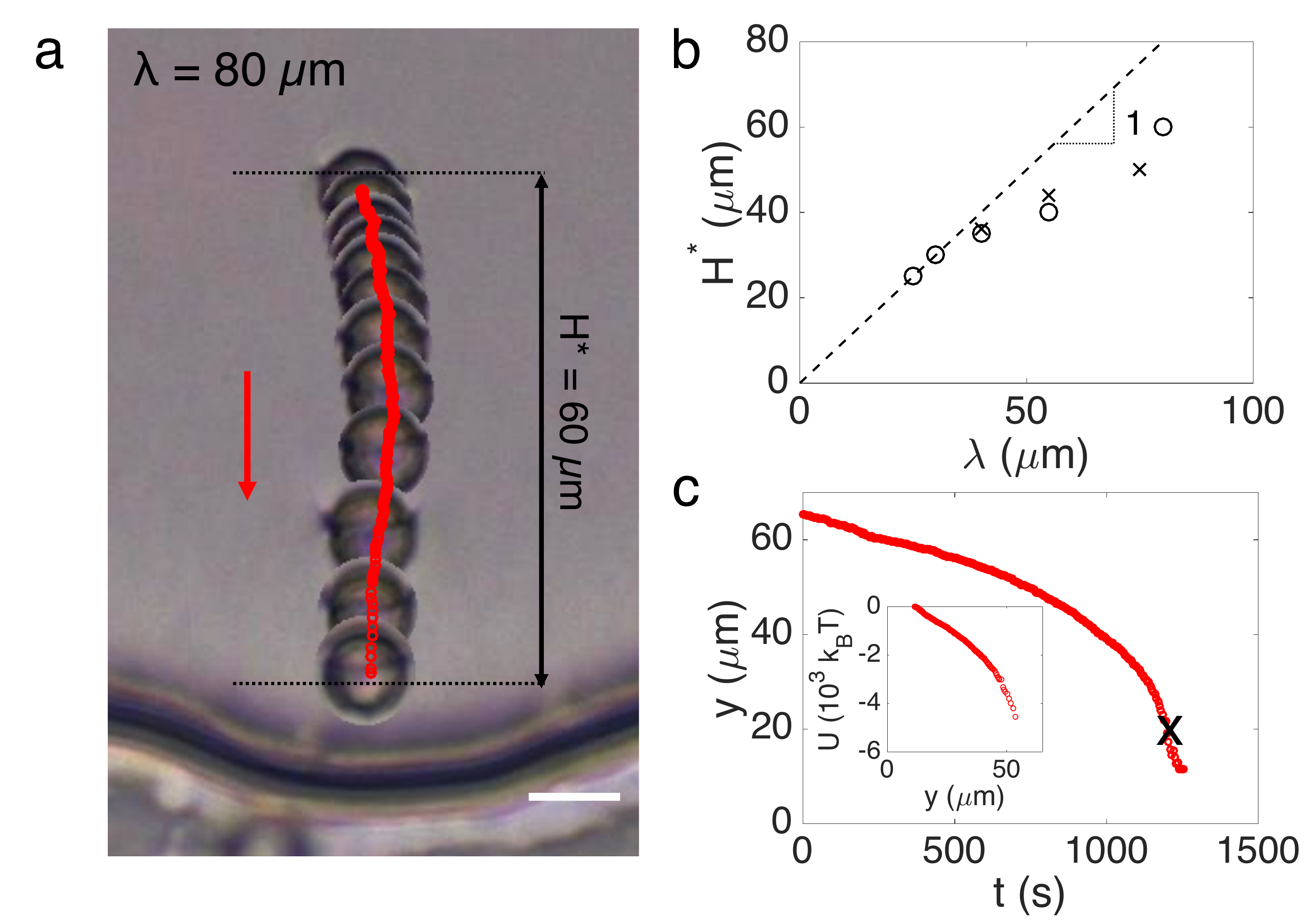}
 \captionof{figure}{{\bf Range of colloid-wall interaction increases with $\lambda$.} A ferromagnetic homeotropic colloid with a Saturn ring defect is used to establish the range of interaction $H^*$ of the colloid with the wall. (a) An equal time step ($\Delta t$ = 125 s) image is shown for the case $\lambda=80\ \mu$m, $H^* = 60\  \mu$m. (b) Range of interaction $H^*$ versus the wavelength of the feature $\lambda$ for homeotropic droplets (open circles) and homeotropic colloids (crosses). (c) The position of the particle $y$ with respect to time $t$. Inset: Energy dissipated to viscosity along a particle trajectory $U$ with respect to the particle position $y$. The cross shows where we truncate the trajectory for integration along the path to infer the dissipation. The scale bar is 10 $\mu$m.}
}

{\bf Equilibrium position.} The wall shape also determines the equilibrium position $H_e$ near a well. In fact, we show that the particles do not always dock very close to the wall, rather, they find stable equilibrium positions at predictable distances from contact with the hills and valleys. We probe this phenomenon by varying colloid radius $a$ and wall radius of curvature $R$ (Fig. 3a). At equilibrium, the location of the center of mass of the colloid $y$= $H_{\mathrm{COM}}$ is equal to $R$. That is, the colloid locates at the center of curvature of the well (Fig. 3b). In this location,the splay of the NLC director field from the colloid matches smoothly to the splay sourced by the circular arc that defines the well. As $R$ increases, this splay matching requirement moves the equilibrium position of the colloid progressively away from the wall. 
However, for wide wells with $R \gg 2a$, the elastic energy from the wall distorts the Saturn ring, displacing it away from the wall (Fig. 3c). 
 When this occurs, colloids equilibrate at loci closer to the wall. For all such colloids, the height of Saturn rings (Fig. 3a crosses: $y$ = $H_{\mathrm{defect}}/a$) and that of the center of mass of the particles (Fig. 3a open circles: $y$ = $H_{\mathrm{COM}}/a$) do not coincide. Specifically, the particle moves closer to the wall and the particle-defect pairs become more dipole-like. For comparison, we plot the center of mass of particles with dipolar defects sitting near the wall (Fig. 3e). We observe that, when the colloid radius is similar to the radius of the wall ($R/a < 2$), there is a similar ``splay-matching" zone for the dipoles; however, as we increase $R/a$, the behavior changes. In this regime, the dipole remains suspended with its hedgehog at a distance of roughly $y$ = $H/a = 3$ from the base of the well for wells of all sizes. The equilibrium distance of particles with distorted Saturn rings (Fig. 3a open circles) is intermediate to equilibria for particles with undistorted Saturn rings and colloids in dipolar configurations with point defects. LdG simulation corroborates the finding that dipoles and quadrupoles equilibrate at different distances from the wall, and that the particles with dipolar defects sit deeper in the well than those with Saturn ring  (Fig. S5a,b).

\begin{figure*}[htb]
 \centering
  \includegraphics[width=2\columnwidth]{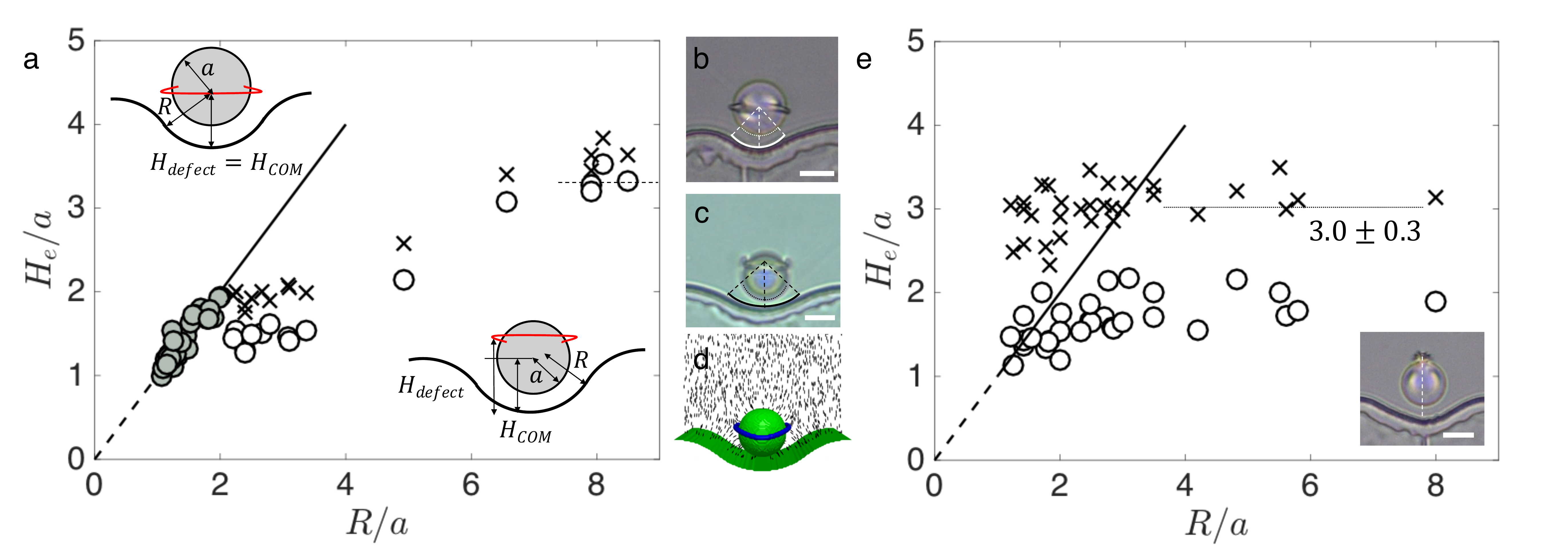}\\
 \caption{{\bf Particle-wavy wall interactions: Splay matching and defect displacement regimes.} (a) Filled grey circles denote splay matching cases, where the Saturn ring sits at the equatorial position ($H_{\mathrm{defect}}/a$ = $H_{\mathrm{COM}}/a$). Crosses denote location of distorted Saturn rings, $H_{\mathrm{defect}}/a > H_{\mathrm{COM}}/a$. Open circles indicate the height of the center of mass (COM) of the colloid. The dotted line denotes flat wall limit. Inset: Schematics of  splay matched and displaced defect cases. Experimental bright field microscopy images of (b) splay matched and (c) defect displacement cases. (d) LdG simulation of the geometry in (c) shows the displacement of the ring. (e) Heights of the center of mass (COM, open circles) and hedgehog defects (crosses) of the colloid with dipole defects. Inset: $a =7.5  \,\mu$m, $R = 17.5  \,\mu$m, $H_{\mathrm{defect}}/a = 3$. The scale bars are 10 $\mu$m.}
\end{figure*}

The heat map in Fig. 1a summarizes the results of LdG simulations for the energy landscape around a microparticle with a Saturn ring defect at various locations in the domain. A colloid positioned directly above a well moves down the steepest energy gradient, which corresponds to a straight path toward the wall. The energy minimum is found when the particle is at a height determined by $R/a$, consistent with our experiments (Fig. 3b). For large $R$, the LdG simulations reveal a distorted Saturn ring around the colloid (Fig. 3d). We also note that at $R/a = 7$, we find $H_{\mathrm{COM}}/a = 3.5$, which corresponds to the equilibrium distance of colloids repelled from a flat wall. However, even at these wide radii, the elastic energy landscape above the undulated wall differs significantly from the repulsive potential above a planar boundary, which decays monotonically with distance from the wall \cite{chernyshuk2011theory}. For colloids above the wide wells, energy gradients in the $y$-direction are small, but gradients in the $x$-direction are not. As a result, particles migrate laterally and position themselves above the center of the wells. These dynamics differ from particles above a planar wall, where the particles diffuse freely parallel to the wall.

{\bf Quadrupole to dipole transition.} For micron-sized colloids in an unbounded medium, the dipole is always the lowest energy state \cite{lubensky1998topological}. Electrical fields \cite{loudet2001application}, magnetic fields \cite{stark1999director} or spatial confinement \cite{gu2000observation} stabilize the Saturn ring configuration. In prior research, a colloid with Saturn ring defect, stabilized by confinement far from the wavy wall, became unstable and transformed into a dipolar structure near the wavy wall \cite{luo2016experimental}. In that work, wells with radii $R$ similar to the colloid radius $a$ were used, and categorized a set of narrow well geometries that prompted this transformation. However, in those settings, the transformation occurred very near the wall, where dynamics of the colloid and surrounding liquid crystal were strongly influenced by the details of wall-particle hydrodynamic interactions and near-wall artifacts in the director field. Here, to avoid these artifacts, we use wells with a smooth boundary where $R > a$ and amplitude $A \sim a $ (specifically, 2$A$ = 15 $\mu$m and $\lambda$ = 60 $\mu$m or 2$A$ = 25 $\mu$m, and $\lambda$ = 100 $\mu$m.). These wells are deeper and are best described as semicircle arcs with rounded corners. 

We exploit these wider wells to position a colloid with a companion Saturn ring several radii above the wall. The elastic energy field distorts the Saturn ring, and drives a gentle transition to a dipolar defect configuration, as shown in Fig. 1c in time lapsed images. The location of the colloid's center of mass (COM) and the evolution of the polar angle of maximum deflection $\theta$ are tracked and reported in Fig. 4a. This transition is not driven by hydrodynamics; the Ericksen number in this system $\mathrm{Er}= 8 \times 10^{-4}$ , a value two orders of magnitude lower than the critical $\mathrm{Er}= 0.25 $ for a transition from quadrupole to dipole driven by flow \cite{khullar2007dynamic}.

Far from the wavy wall, the effect of the two parallel walls that confine the colloid is similar to that of an external electromagnetic field or to a weakening of the anchoring on the surface of the colloid \cite{stark}, all of which make the Saturn ring configuration either stable or metastable \cite{stark1999director}. The wavy wall, however, exerts an asymmetrical elastic stress on the Saturn ring, displaces it away from that wall, and ultimately destabilizes this configuration. We have performed an experiment in which we allowed the Saturn ring to transition to a dipole near the wall, and then rapidly removed the elastic stress by driving the particle away using a magnetic field (Supplemental Video V1). The dipole remained stable, which indicates that, under our experimental condition, the dipole is the stable state and the Saturn ring is metastable. We can consider the polar angle $\theta$ and the director field as our ``reaction coordinate" to characterize the transition between the Saturn ring state ($\theta=\pi/2$) and the dipolar state ($\theta=\pi$). We assume that the maximum of the energy barrier between these two states far from the wall will be found at an intermediate angle $\theta_B$ (Fig. S5c). The elastic energy from the wall lowers the energy barrier to the transformation, allowing it to occur.

Previously, Loudet and collaborators \cite{loudet2002line} studied the transition of a colloid with a Saturn ring defect to a dipolar configuration in an unbounded medium. In that study, the Saturn ring configuration was stabilized by an electric field, and the transition occurred after the electrical field was removed. The process took place over the course of 10-60 seconds and was found to be independent of the colloid size. Although these two sets of experiments take place in very different physical systems (confined vs.\ unconfined, withdrawal of an electric field versus an applied stress field via boundary curvature), the slow initial dynamics and the total time of transition are common features shared by both. Our results bear remarkable similarities with the dynamics shown in Ref. \cite{loudet2002line} (Fig. 4b-c). In our system, the Saturn ring is metastable, the stabilization is provided by confinement from parallel glass, and destabilized by elastic stress from the wavy boundary.

{ \centering
 \includegraphics[width=\columnwidth]{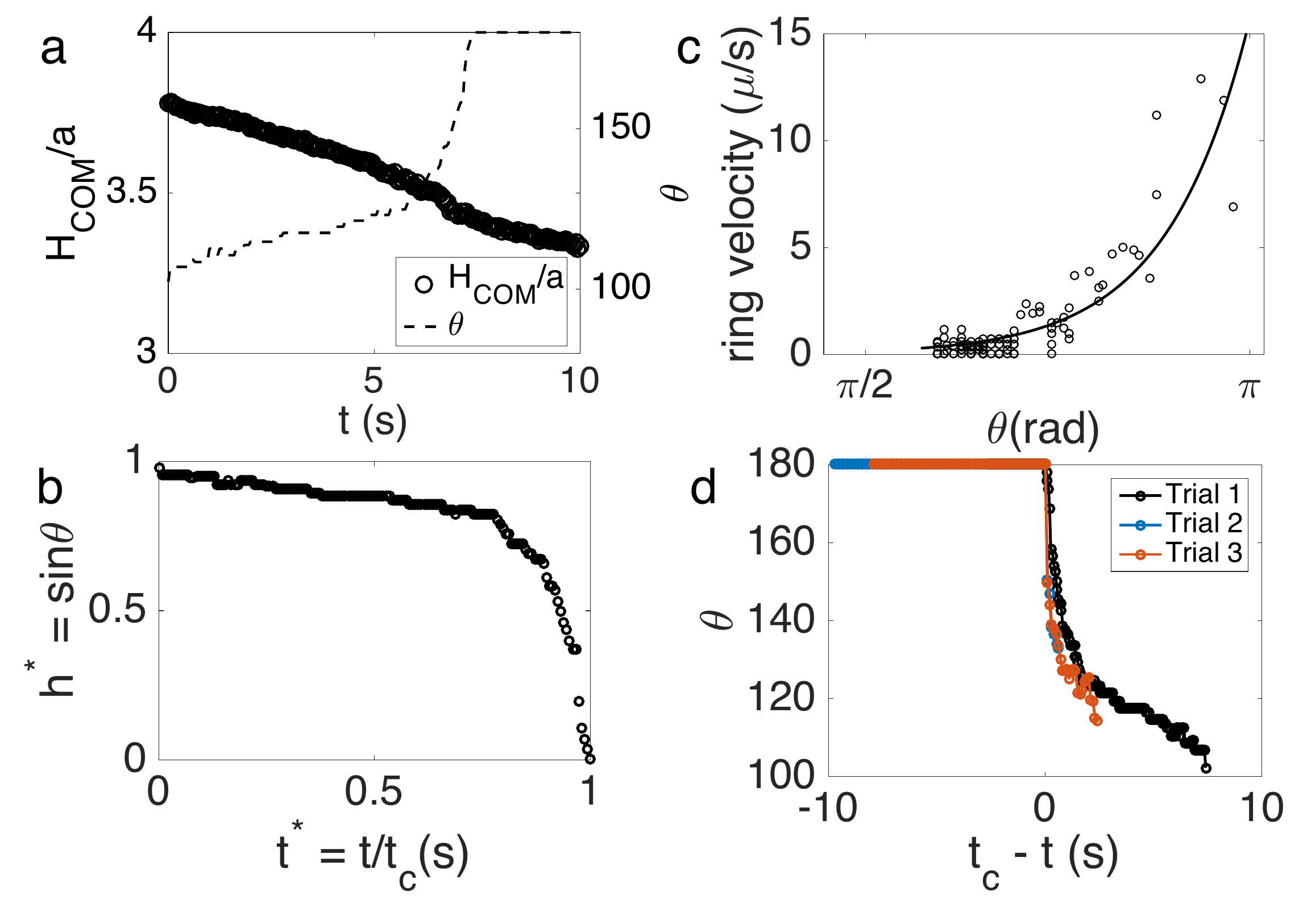}\\
 \captionof{figure}{{\bf Dynamics of the quadrupole to dipole transition.} (a) Tracking center of mass (COM) and polar angle $\theta$ evolution during the quadrupole to dipole transition. Initially, the colloids assume the $\theta = 90^\circ$ (dipolar) configuration, which gradually evolves to $\theta = 180^\circ$ as the COM continuously moves towards the wall. After the transition has taken place, the COM continues to approach the wall. (b, c) Reduced ring size and velocity from our system reveal similar dynamics of transition as shown in Fig. 2 in \cite{loudet2002line}. (d) $\theta$ vs. $t_c-t$ plot shows three experimental runs of transition in similar geometry. In (b-d), $t_c$ is the time at which $\theta=90^\circ$.}
}

Further experiments reveal that we are able to exert control over the transition by controlling the shape of the wells. In this case, we made deep wells of either $2A = 15 \, \mu$m or $2A = 25 \, \mu$m. Then we plot the angle versus $t_c - t$, where $t_c$ is the time $\theta$ reaches $\pi/2$. The $\theta$ variations for 3 cases for wells of  $2A = 15 \, \mu$m (Fig. 4d) superpose. We show additional trials in Fig. S6; the dynamics are reproducible across samples of different sizes, even in the case where debris are collected by the topological defects on the way. While Loudet et al. observed a propulsive motion opposite to the defect motion attributed to back flow from reorientation of director field, in our system the motion is smooth and continuous as the colloid passes through the spatially varying director field. However, the velocity of the droplet decreases right after transition; we attribute this to the change in the drag environment (Fig. 4a and Fig. S6b)

In shallow wells ($A < a$) with small radii of curvature ($R \sim a$), the particle docks via the familiar lock-and-key mechanism. However, if the radius is large ($R > a$), the well exerts an elastic stress on the director field around the colloid and the Saturn ring remains in the distorted state. The polar angle (Fig. S5c) then ranges from $\theta$ = $103 ^\circ$ to $130 ^\circ$ (maximum deflection). The energy barrier between the Saturn ring configuration and the dipolar configuration persists. However,at a critical angle $\theta_c$, the favorable energy from bend and splay matching eliminate the energy barrier between dipole and quadrupole, allowing completion of the quadrupole to dipole transition. This critical angle of transformation is relatively independent of the colloid size and mode of confinement, and found to be around $\theta_c = 150 ^\circ$ in our experiments, which differs from the one found by \cite{vskarabot2008interactions}. The initial dynamics of the quadrupole to dipole transition is slow, but as $\theta$ increases, so does $d\theta/dt$ (Fig. S5c).

\begin{figure*}[htb]
 \centering
  \includegraphics[width=1.75\columnwidth]{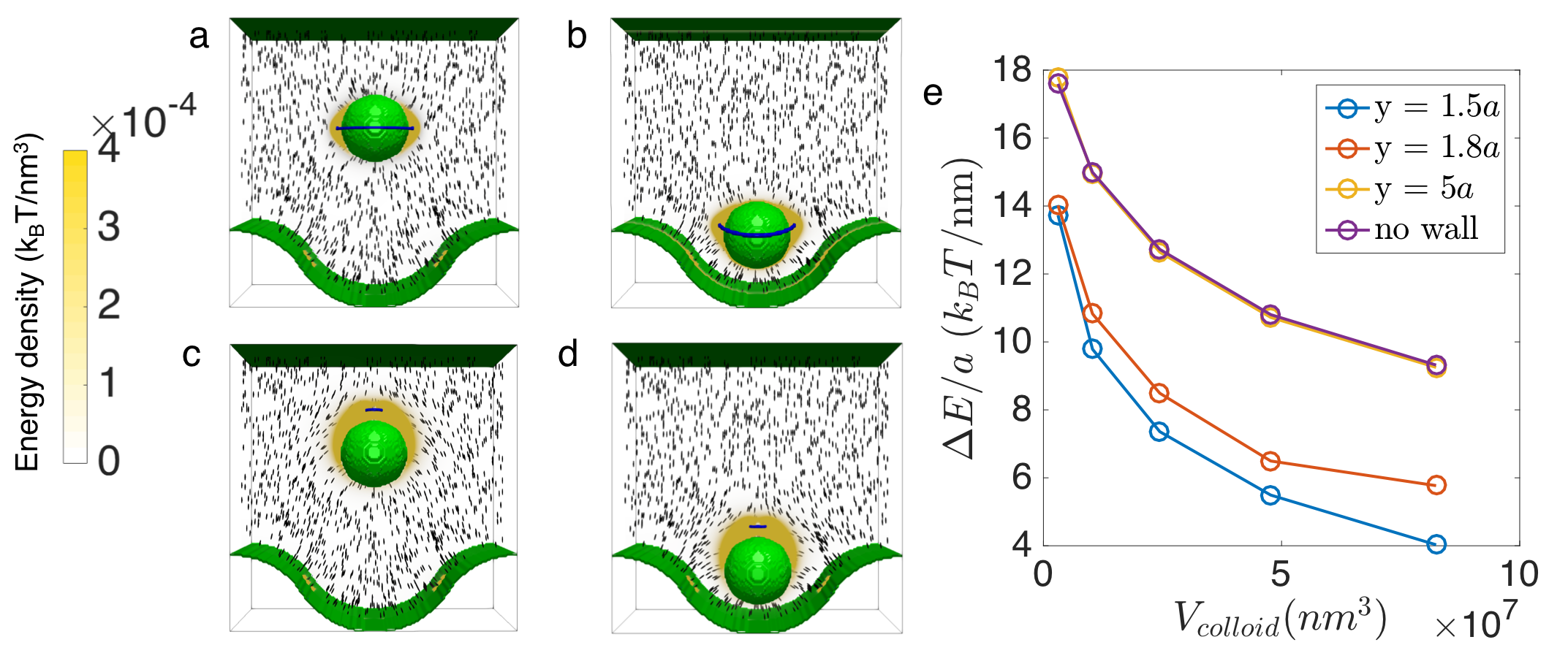}\\
 \caption{{\bf LdG simulation of the energy density for dipole and quadrupole near a wavy boundary.} The numerical energy minimization is performed for various positions of dipole and Saturn ring, with the colloid size and wavy wall geometry held fixed, to find the location of minimum energy. (a) A Saturn ring located at reference state (State 1, $y=5a$) with E = 7289.70 $k_B T$. (b) A Saturn ring located at near wall (State 2, $y = 1.8a$) with E = 7086.20 $k_B T$, a decrease of 203.5 $k_B T$ from State 1. (c) A dipole located at reference state (State 1, $y=5a$) with E = 9041.13 $k_B T$. (d)  A dipole located at near wall (State 2, $y=1.5a$) with E = 8456.12 $k_B T$, a decrease of 585.01 $k_B T$ from State 1. (e) The energy of the dipole and quadrupole are calculated for systems of different size ($a =90, 135, 180, 225, 270\,$ nm, the simulation box and the wall are scaled accordingly). The energy difference between quadrupole and dipole ($\Delta E = E_{dipole} - E_{Saturn \, ring}$) for systems of different sizes at $y = 1.5a$ (blue curve), $y = 1.8a$ (orange curve), $y = 5a$ (yellow curve), and no wavy wall (purple curve).}
\end{figure*}

In deeper wells ($A > a$), the polar angle increases as the colloid migrates into the well. LdG simulation reveals that, in the dipolar configuration, there is less distortion in the director field near the colloid owing to bend and splay matching, and that it is indeed more favorable for a colloid with dipolar defect to locate deep within the well (Fig. 5).  We can compute the energy of a colloid both far (State 1: $y=5a$, reference state) and near the wavy wall (Fig. 5a-d) for both Saturn ring and dipolar configurations (State 2: $y = 1.8a$ and $y = 1.5a$, for Saturn ring and dipolar configuration, respectively). Using identical parameters for the LdG numerics, we can stabilize a dipolar configuration by initializing the director field by the dipolar far-field ansatz \cite{stark2001physics}.  While colloids in both configurations decrease their energy upon moving toward the wall from State 1 to State 2, the decrease in energy is $2.9$ times greater for the dipolar case (Fig. 5a). This change is determined by differences in the gradient free energy, corresponding to reduced distortion in the nematic director field.    

Stark \cite{stark} argues that the stabilization of a Saturn ring under confinement occurs when the region of distortion becomes comparable to or smaller than that of a dipole, assuming the same defect energy and energy density. Yet this argument does not apply here because the presence of the wavy wall strongly alters the energy density at various regions (Fig. 5). Due to limitations in computational power, we cannot model colloids of our experimental scale. This limits our simulations to settings in which the dipole is more energetically costly than the Saturn ring configuration.  However, as we increase the size of the simulation ($a$ = 90, 135, 180, 225, 270 nm), the energy difference between dipole and quadrupole decreases for colloids located at $y = 1.5a$ (Fig. 5e, blue curve), suggesting that at larger system sizes, the dipole may become the stable configuration, in agreement with experiment. The energy difference $\Delta E\, ( = E_{dipole} - E_{Saturn \, ring}$) decreases going from $y = 5a$ (Fig. 5e, yellow curve) to $y = 1.8a$ (Fig. 5e, orange curve), as expected. Note that $\Delta E$ at $y = 5a$ agrees to within 1.15\% with a simulation of colloids in a sandwich cell with no wavy wall (Fig. 5e, purple curve), serving as a valid reference state. Furthermore, we note that the energy difference between a dipole and quadrupole configuration decreases as colloids move closer to the bottom of the well. These results show that the distortion field exerted by the wavy boundary can be considered as a gentle external field, in analogy to electrical, magnetic or flow fields. 

\section*{Discussion, multistable states}

{\bf Multiple paths diverging from unstable points.}
In the preceding discussions, we have focused on attractive particle-wall interactions and associated stable or metastable equilibria. However, the location directly above a hill is an unstable point. When colloids are placed nearby using an external magnetic field, they can follow multiple diverging  paths upon removal of the magnetic field.  The particular paths followed by the colloid depend on small perturbations from the unstable point. 

For example, amongst 28 such trials using an isolated homeotropic colloid with a Saturn ring, a colloid moved along a curvilinear path to the well on its left 11 times, to the well on its right 10 times and was repelled away from the peak until it was approximately one wavelength away from the wall 7 times.  Three sample trajectories are shown in Supplemental Video V2a-c. These trajectories are also consistent with the heat map in Fig. 1a, computed by taking a fixed step size in the direction of the local force as defined by the local energy gradient. The numerically calculated trajectories, and their extreme sensitivity to initial position, are in qualitative agreement with our experimental results. Thus, small perturbation in colloid location can be used to select among the multiple paths. 

These features can be used to launch the colloid from one location to another, propelled by the elastic energy field. To demonstrate this concept, we arranged two wavy walls parallel to each other with the periodic structures in phase, i.e. the hills on one wall faced valleys on the other (Fig. 1f). For wall-to-wall separations more than $2 \lambda$, colloids docked, as expected (Fig. 1f).  For wall-to-wall separations less than $2 \lambda$, a colloid, placed with a magnetic field above the peak on one wall was guided by the NLC elastic energy to dock in the valley on the opposite wall (Fig. 1g), thus effectively extending its range of interaction with the second wall (Supplemental Video V3). In the context of micro-robotics, such embedded force fields could be exploited to plan paths for particles to move from one configuration to another, guided by a combination of external magnetic fields and NLC-director field gradients. 

{\bf Path-planning for colloids with different defect structures.} 

We can tailor unstable points and attractors for these particles, and find important differences between the behavior of colloids attracted to wells and those attracted to hills. For example, a dipole pointing away from the wall (Fig. 6a) behaves like a colloid with companion Saturn rings in several ways.  Both are attracted over a long range to equilibrate in wells, and both have unstable points above hills. Also, when released from this unstable point, both defect structures can travel in three distinct directions (left, right and away from the wall, Fig. 6a).
On the other hand, dipoles pointing toward the wall (Fig. 6b) behave differently. They are attracted to stable equilibria near hills, and are unstable near wells. Interestingly, when released from a point near a well, these colloids can travel only toward one of the adjacent hills. That is, there is no trajectory above the well that drives them in straight paths away from the wall. Colloids with planar anchoring form boojum structures which behave similarly (Fig. 6c); they equilibrate near the hills, and follow only two sets of possible paths when released from unstable points above a well. The ability to drive particle motion with a gently undulating wall is thus not limited to colloids with companion Saturn rings; the wall also directs the paths of dipolar colloids with homeotropic anchoring and colloids with planar anchoring, decorated with boojums. The interaction ranges for various colloid-defect configurations are summarized in Fig. 6d; while colloids with each defect structure have distinct equilibrium distances from a flat wall (Fig. S7), the range of interactions follow similar trends as functions of $\lambda$ (Fig. 6d). 

{ \centering
 \includegraphics[width=\columnwidth]{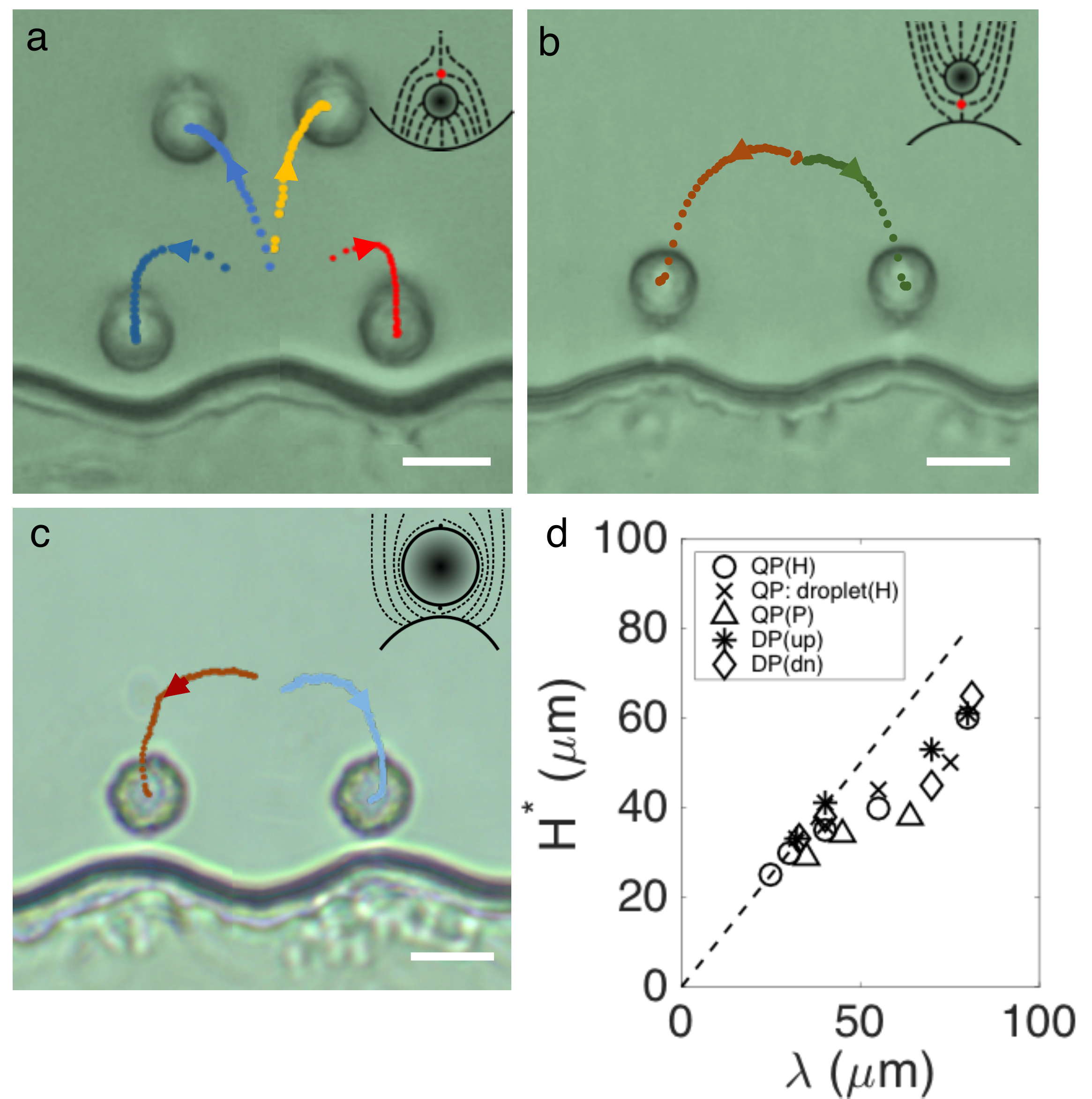}\\
 \captionof{figure}{{\bf Multiple states and reconfigurable docking.} Particle paths are illustrated by points that indicate particle COM position over time; time step $\Delta t$=5 seconds between neighboring points. The colored dots denote (a) 4 representatives trajectories (out of 12) of an upward-oriented dipole, (b) 2 representative trajectories (out of 11) of a downward-oriented dipole, and (c) 2 representative trajectories (out of 14) of a planar-anchoring colloid with two boojums released between two neighboring wells. Inset: A sketch of the director field around (a) an upward-oriented dipole docked inside the well, (b) a downward oriented dipole and (c) a planar-anchoring colloid perched on top of a hill. The scale bars are 10 $\mu$m. (d) The range of interaction $H^*$ as a function of $\lambda$ is similar for homeotropic (H) and planar (P) anchoring, for hedgehog (DP) and Saturn ring (QP) defects, and for solid colloids and droplets. }
}

These results indicate that the range of repulsion differs for hills and wells. This is likely related to the differences in the nematic director field near these boundaries. While converging splay field lines are sourced from the well, divergent splay field lines emanate from the hill. Both fields must merge with the oriented planar anchoring far from the wall. As a result, hills screen wells better than wells screen hills.

We can exploit these wall-dipole interactions to shuttle the colloid between parallel walls. For walls positioned with their wavy patterns out-of-phase (Fig. 7, Supplemental Video V4), dipoles with point defect oriented upwards are repelled from initial positions above hills on the lower wall and dock on the hill on the opposite wall. However, for walls with their patterns in phase, dipoles with defects oriented downwards released from an initial position above a well dock either at an adjacent hill on the same wall (Fig. 7c, Supplemental V5a), or in an attractive well on the opposite wall (Fig. 7d, Supplemental Video V5b).

{ \centering
 \includegraphics[width=0.9\columnwidth]{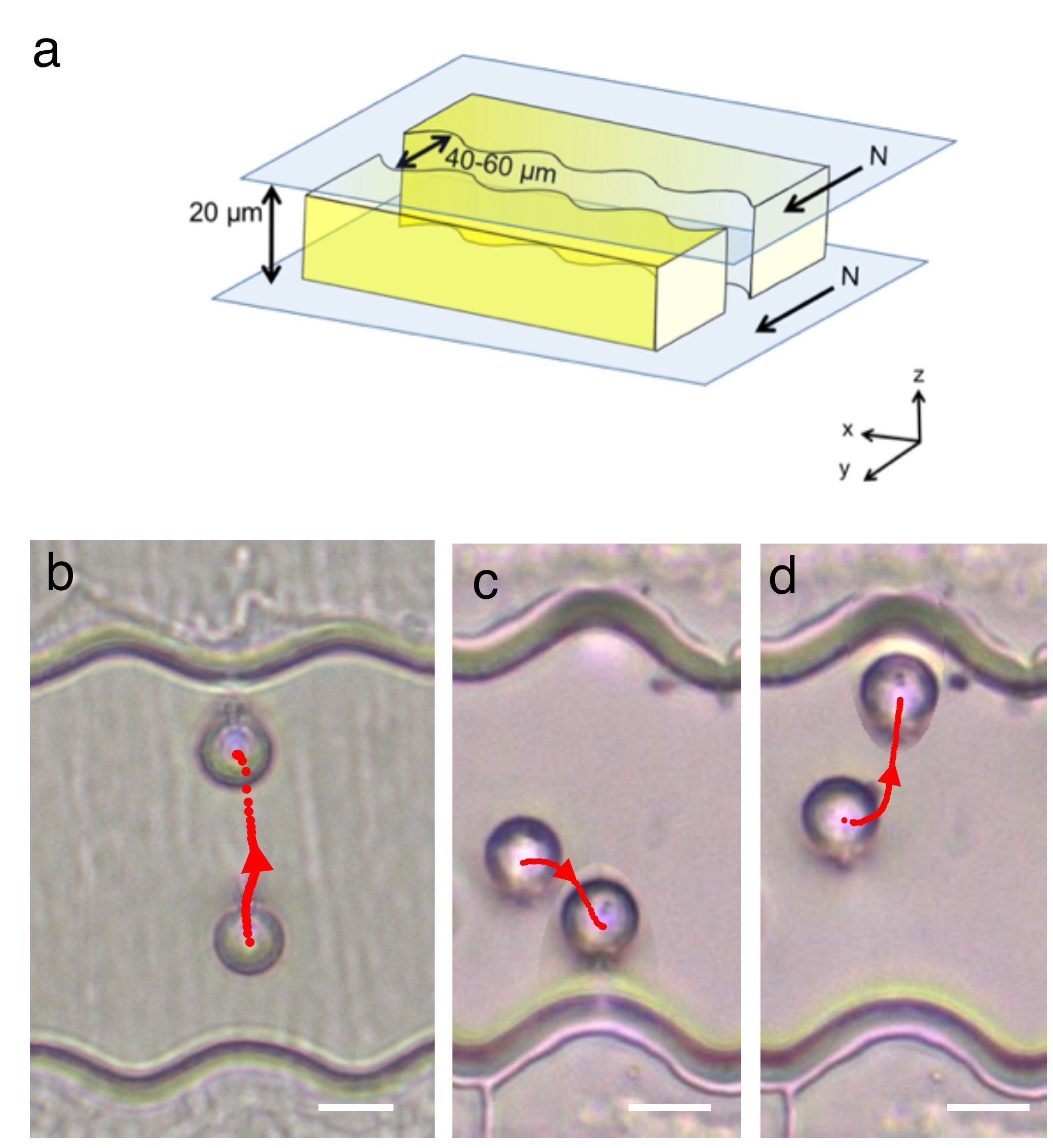}\\
 \captionof{figure}{{\bf Repulsion and bistable docking of dipoles.} (a) Schematics of two parallel walls with a gap comparable to $\lambda$ between them. The waves of the wall are either out of phase with hill to hill configuration such as in (b) or in phase with hill to valley such as in (c-d). The scale bars are 10 $\mu$m.}
}

Finally, we demonstrate that particles moving in weak flow can select preferred docking sites along the wavy wall.  Wells of different wavelengths create energy gradients that decay at distinct rates. Placing wells of different wavelength adjacent to each other offers additional opportunities for path planning. In one setting, a colloid can sample multiple wells of varying sizes under a background flow in the $x$ direction (Fig.\ 8). The outcome of whether the colloid docks or continues to be advected is determined by a balance between viscous forces that drive $x$-directed motion and attractive and repulsive interactions with the wall. In a separate experiment, we place tracer particles in the background while a sampling/docking event takes place (Fig. S8). The tracer particle travels along a straight path while the colloid near the wall follows a more complex trajectory, eventually docking in a well that, as for Goldilocks, protagonist of a beloved children story, is ``just right". The complexity of the colloid's path confirms that the elastic energy field plays an important role in guiding the motion of the colloid to its preferred well. Such interactions open interesting avenues for future studies, in which the rates of motion owing to elastic forces and those owing to applied flows are tuned, and the trapping energy of the docking sites are tailored, e.g. for colloidal capture and release.   


{ \centering
 \includegraphics[width=\columnwidth]{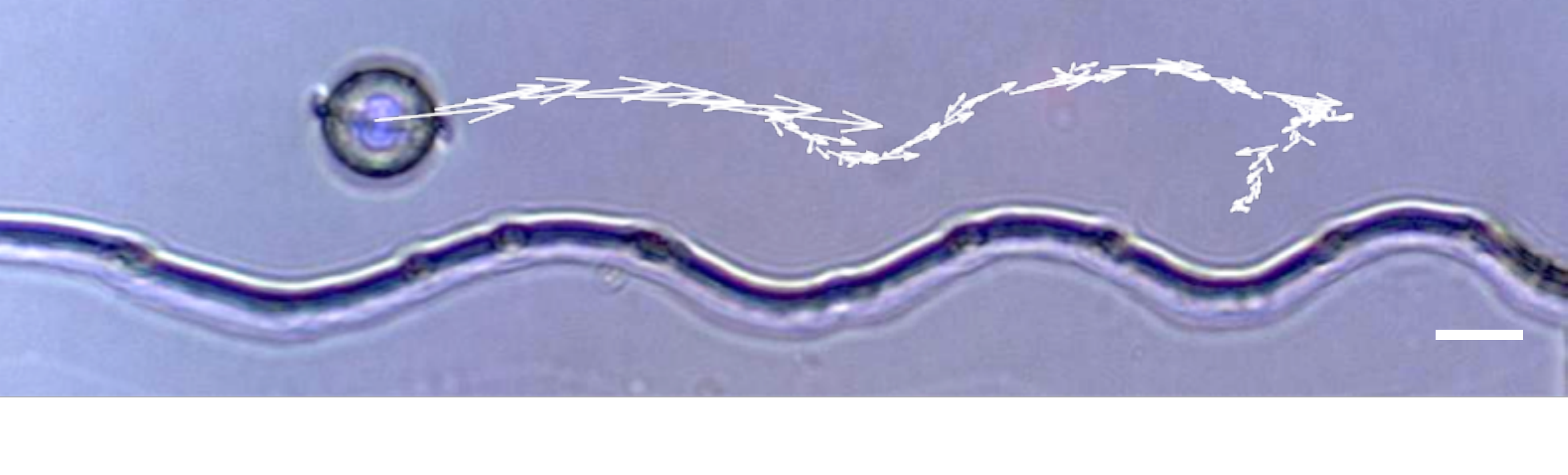}\\
 \captionof{figure}{{\bf ``Goldilocks'': Colloid docks in a preferred well.} A particle finds the lowest energy locations under a biasing flow (Supplemental Video V6), ending in the well that best matches its curvature. The length of the arrow is proportional to the instantaneous velocity. The scale bar is 10 $\mu$m.}
}

\section*{Conclusions} 
The development of robust methods to drive microscopic objects along well defined trajectories will pave new routes for materials assembly, path planning in microrobotics and other reconfigurable micro-systems. Strategies developed within NLCs are one means to address these needs.  Since the strategies developed in liquid crystals depend on topology, confinement, and surface anchoring, which can be manipulated by changing surface chemistry or texture on colloids with very different material properties, they are broadly applicable across materials platforms. We have developed controllable elastic energy fields in NLCs near wavy walls as a tool to manipulate the ranges of attraction and to define stable equilibiria. We have also exploited elastic energy fields to drive transitions in topological defect configurations. The near-field interaction between the colloid and the wall rearranges the defect structure, driving a transition from the metastable Saturn ring configuration to the globally stable dipolar configuration for homeotropic colloids. 

We account for this transformation by means of an analogy between confinement and an external magnetic field. As these defect sites are of interest for molecular and nanomaterials assembly, the ability to control their size and displacement will provide an important tool to improve understanding of their physico-chemical behavior, and potentially to harvest hierarchical structures formed within them.

Furthermore, we have developed the concept of repulsion from unstable points as a means to dictate paths for colloids immersed within the NLCs. We have identified unstable sites from which multiple trajectories can emerge, and have used these trajectories to propel particles, demonstrating the multistability made possible by the wavy wall geometry.

\section*{Methods}
{\bf Assembly of the cell.} We have developed a wavy wall confined between two parallel plates as a tool to direct colloid assembly. The wavy wall is configured as a bounding edge to the planar cell. The NLC cell and the walls were fabricated following the procedure in \cite{luo2016experimental}. The procedure is briefly outlined here. The wavy walls are made with standard lithographic methods of SU-8 epoxy resin (MicroChem Corp.). The walls have period $\lambda$ ranging from $27-80$ $\mu$m and consist of smoothly connected circular arcs of radius $R$ between $7-40$ $\mu$m. These strips, of thickness between $20-28$ $\mu$m, are coated with silica using silica tetrachloride via chemical vapor deposition, then treated with DMOAP (dimethyloctadecyl[3-(trimethoxysilyl)propyl]). The wavy wall is sandwiched between two antiparallel glass cover slips, treated with 1\% PVA (poly(vinyl alcohol)), annealed at 80 $^{\circ}$C for one hour and rubbed to have uniform planar anchoring. Once assembled, the long axis of the wall is perpendicular to the oriented planar anchoring on the bounding surfaces. We observed that in some LC cells the actual thickness was larger than expected, which we attribute to a gap above the strip. In those cases we noticed that some small colloids could remain trapped between the wavy strip and the top glass surface, so the effective thickness could be as large as $35-40$ $\mu$m. 

{\bf Particle treatment and solution preparation.} We use the nematic liquid crystal 5CB (4-cyano-4'-pentylbiphenyl, Kingston Chemicals) as purchased. We disperse three types of colloids in the 5CB. The size and polydispersity of the colloids are characterized by measuring a number of colloids using the program FIJI. (1) $a = 7.6 \pm 0.8$ $\mu$m silica particles (Corpuscular Inc.), treated with DMOAP to have homeotropic anchoring. (2) $a = 4.3 \pm 0.4$ $\mu$m ferromagnetic particles with polystyrene core and coated with chrome dioxide (Spherotech, Inc.), treated with DMOAP, an amphiphile that imposes homeotropic anchoring, and with PVA for planar anchoring. (3) $a = 4.3-8$  $\mu$m custom-made emulsion droplets where water phase was loaded with magnetic nanoparticles and crosslinked. The oil phase consisted of 5CB mixed with 2wt\% Span 80. The water consisted of a 50:50 mixture of water loaded with iron oxide nanoparaticle and a pre-mixed crosslinking mixture. The magnetic nanopowder iron (II, III) oxide (50-100 nm) was first treated with citric to make it hydrophilic. The crosslinking mixture was pre-mixed with HEMA (2-hydroxyl ethyl methacrylate): PEG-DA (poly(ethylene glycol) diacrylate): HMP (2-hydroxyl-2-methylpropiophenone) in 5:4:1 ratio. Water and oil phases were emulsified with a Vortex mixer to reach desired colloid size range. The two were combined in a vial treated with OTS (trichloro(octadecyl)silane) to minimize wetting of the wall by the water phase during the crosslinking process. All chemicals were purchased from Sigma Aldrich unless otherwise specified. The emulsion was crosslinked by a handheld UV lamp (UVP, LLC) at  = 270 nm at roughly P = 1mW/cm$^2$ for 3 hours. The emulsion was stored in a refrigerator for stability. Span 80 ensured that the liquid crystal-water interface would have homeotropic anchoring. The magnetic droplets are very polydispersed due to the emulsification process. However, when we compare their behavior with the silica and feromagnetic colloids, we only compare colloids and droplets of similar sizes.

{\bf Imaging.} The cells form a quasi-2D system that is viewed from above. In this view, the wavy wall is in the plane of observation. The homeotropic colloids dispersed in the NLC are located between the top and bottom coverslips. These colloids are levitated away from both top and bottom surfaces by elastic repulsion \cite{pishnyak2007levitation}. The cell was imaged with an upright microscope (Zeiss AxioImager M1m) under magnification ranging from 20x to 50x. The dynamics of the colloid near the wavy wall are recorded in real time using optical microscopy. Additional information regarding the  director field configuration is also gleaned using polarized optical microscopy (POM).  

{\bf Application of a magnetic field.} The magnetic field was applied by using a series of 8 NdFeB magnets (K\&J Magnetics, Inc.) attached to the end of a stick. The magnets was placed roughly 0.5 cm from the sample; the field applied is estimated to be roughly 40-60 mT, far below the strength required to reorient the NLC molecules, but sufficiently strong to overcome the drag and move magnetic droplets and particle in arbitrary directions. 

{\bf Numerical modeling by Landau-de Gennes (LdG) simulation.} Numerical modeling provides insight into the NLC director field in our confining geometries.  We use the standard numerical Landau-de Gennes (Q-tensor) approach \cite{ravnik2009landau} with a finite difference scheme on a regular cubic mesh. This approach is widely used to compute regions of order and disorder in bounded geometries through a global free energy minimization. The Q-tensor is a second-rank, traceless, symmetric tensor whose largest eigenvalue is the order parameter $S$ in the NLC. Using the Landau-de Gennes approach, at equilibrium, the 3-D director field and the locations of defect structures for a given geometry are predicted. The nematic director field, a headless vector field (i.e.  $-\bf{n} \equiv \bf{n}$), represents the average direction of an ensemble of molecules of size comparable to the correlation length at any point in the system. The geometry of the system, the boundary conditions, and elastic constants for the NLC are inputs to the numerical relaxation procedure. Specifically, the particle is bounded by walls with oriented planar anchoring separated by thickness $T = 4a$. The anchoring at the boundary opposite of the wavy wall is set to zero, and that of the flat plates sandwiching the colloid and the wavy wall is set to oriented planar. The Nobili-Durand anchoring potential is used \cite{nobili1992disorientation}. Defects are defined as the regions where the order parameter $S$ is significantly less than than the bulk value. The mesh size in our simulation is related to the correlation length in the NLC, and corresponds to 4.5 nm. Due to the difference in scale, the exact final configurations of numerics and experiment must be compared with caution; nevertheless, it is an invaluable tool to corroborate and elucidate experimental findings. Because the size of simulation is much smaller, much stronger anchoring is applied. For most of our results, infinite anchoring strength is applied unless otherwise specified. To simulate dipoles, we vary the material constants $B$ and $C$ so that the core energy of the defect is 2.6x higher to compensate for the small system (details can be found in Supplemental Materials). In addition, we also use an initial condition with a dipolar configuration about the colloid: ${\bf n}({\bf r}) = \hat i + PR_c^2\frac{{\bf r}-{\bf r_c}}{|{\bf r}-{\bf r_c}|^3}$, where $R_c$ is the colloid radius, ${\bf r_c}$ is the location of the colloid center, $P=3.08$ is the dipole moment, and $\hat i$ is the far-field director \cite{stark2001physics}. This expression is applied only in a sphere of radius 2$R_c$ around ${\bf r_c}$.

{\bf Numerical modeling by COMSOL.} To describe some aspects of the director field in the domain (Fig. S1b), we employ the  common simplification in nematic liquid crystal modeling known as the one-constant approximation: $K_1=K_2=K_3\equiv K$.  If there is no bulk topological defect, then the director field is a solution to Laplace's equation $\nabla^2 {\bf n} = 0$, which can be solved by COMSOL separately for the two components $n_x$ and $n_z$, from which $n_y$ is obtained by the unit length restriction on $\mathbf{n}$. In COMSOL, this is easiest implemented by the Electrostatics Module. The model solves the equivalent electrostatic problem of $\nabla^2 V = 0$, which gives us $n_x$ and $n_z$. Customized geometry, such as the wavy wall, can be built with the geometry builder. We mesh the space with a triangular mesh and calculate the director field components; the results are then exported in grid form and post-processed in MATLAB. 

\section*{Acknowledgement}
This work is supported by the Army Research Office, Grant W911NF1610288. We thank Dr. Sarah Hann for treatment of iron oxide NPs, Dr. Laura Bradley for useful discussion on synthesizing magnetic droplets, Prof. Ani Hsieh, Dr. Shibabrat Naik, Dr. Denise Wong, and Dr. Edward Steager for useful discussion on magnetic control and path-planning. 

\section*{Author contributions} 
K.J.S, F.S., and Y.L. designed the project. Y.L. performed research. Y.L. and D.A.B. performed numerical modeling and theoretical analysis. K.J.S, F.S., Y.L. and D.A.B. wrote the manuscript.

{\bf Competing interests:} The authors declare no competing financial interests.

\end{multicols}

\end{document}